\documentclass[12pt,english]{article}
\usepackage[latin9]{inputenc}
\usepackage[letterpaper]{geometry}
\usepackage{xcolor,graphicx}
\usepackage[american]{babel}
\usepackage{array}
\usepackage{float}
\usepackage{multirow}
\usepackage{amsmath,amssymb,amsthm,mathrsfs}
\usepackage{setspace}
\usepackage{esint}
\usepackage{pdfsync} 
\usepackage{booktabs}
\usepackage[compact]{titlesec}
\usepackage[authoryear]{natbib}
\PassOptionsToPackage{normalem}{ulem}
\usepackage{ulem}
\usepackage[unicode=true,
bookmarks=true, bookmarksnumbered=false,bookmarksopen=false,
breaklinks=false,pdfborder={0 0 1},backref=false,colorlinks=false]
{hyperref}
\usepackage{breakurl}
\bibpunct{(}{)}{;}{a}{}{,}

\begin{document}
\title{\bf Inference about ATE from  Observational Studies with Continuous
  Outcome and Unmeasured Confounding}

\author{TAO LIU$^\ast$,~~ JOSEPH W.\ HOGAN
  \\[4pt]
  \textit{Department   of   Biostatistics,  Center   for   Statistical
    Sciences,}\\
  \textit{Brown University School  of Public Health, }\\\textit{Box GS-121-7,
    Providence, RI 02912, U.S.A. }
  \\[2pt]
  {$^\ast$tliu@stat.brown.edu}
}


\maketitle


\begin{abstract} {For  settings with a  binary treatment and  a binary
    outcome, instrumental variables can be used to construct bounds on
    a causal  treatment effect.  With continuous  outcomes, meaningful
    bounds  are more  difficult to  obtain because  the domain  of the
    outcome is typically  unrestricted.  In this paper,  we combine an
    instrumental variable and subjective assumptions in the context of
    an observational  cohort study of HIV-infected  women to construct
    meaningful   bounds  on   the  initial-stage   causal  effect   of
    antiretroviral therapy  on CD4 count.  The  subjective assumptions
    are encoded in terms of the potential outcomes that are identified
    by observed data as well  as a sensitivity parameter that captures
    the  impact of  unmeasured confounding.   Measured confounding  is
    adjusted using the method  of inverse probability weighting (IPW).
    With extra  information from  an IV, we  quantify both  the causal
    treatment effect and the degree of the unmeasured confounding.  We
    demonstrate our method by analyzing data from the HIV Epidemiology
    Research Study. }
\end{abstract}

\vfill

\global\long\def\propscore{e} 
\global\long\def\MIPW{\mathrm{MIPW}}
\global\long\def\AT{{\mathcal{P}_{11}}}
\global\long\def\CM{{\mathcal{P}_{01}}}
\global\long\def\NT{{\mathcal{P}_{00}}}
\global\long\def\DE{{\mathcal{P}_{10}}} 
\global\long\def\BB{b}
\global\long\def\TT{\tau}
\global\long\def\consis{\stackrel{p}{\longrightarrow}}
\global\long\def\hyph{\mbox{-}} 
\global\long\def\logit{\textrm{logit}}

\section{Introduction}

Observational  studies offer  an important  alternative to  randomized
clinical  trials  when assigning  a  treatment  to study  subjects  is
unethical or  practically impossible  \citep{Rosenbaum:2002}.  However
analyzing data from such studies often confronts the difficulty that a
direct comparison between the treated  and untreated subjects does not
necessarily  show   the  causal  effect   of  the  treatment   due  to
confounding.  Informally,  confounding is caused by  factors that have
causal  effects on  both  treatment and  outcome \citep[see][for  more
rigorous    discussions   and    definitions   of    confounding   and
confounders]{VanderWeele:2011}.  To adjust for the confounding effect,
observational studies typically include collecting a set of covariates
with  the  hope  that  most  confounders  if  not  all  are  measured.
Statistical methods  for controlling for the  confounding effect under
the  assumption  of  no unmeasured  confounding  include  multivariate
adjustment via  regression models,  propensity score  risk adjustment,
propensity  score matching,  and inverse  probability weighting  (IPW)
\citep[][among others]{Bang2005, Kang2007, Robins1994, Rosenbaum:1984,
  Hogan2004, DAgostino:1998, Robins2000}.

The assumption  of no unmeasured confounding is  untestable in general
and very often implausible.  In  this case, making causal inference on
the treatment effect as well as obtaining a quantitative measure about
the  degree   of  unmeasured  confounding  is   imperative.   The  two
objectives  are generally  not  achievable in  a single  observational
study, but  possible given the  existence of an  instrumental variable
(IV).

The  IV  methods  can  be  traced back  to  1920's  \citep{Stock:2003,
  Wright:1928}, and  have been extensively  implemented in econometric
and  recent bio-medical  research.   Loosely speaking,  an  IV can  be
envisioned as  a `randomizer' which  varies exogenously, has  a causal
effect on treatment received, but  has no causal effect on the outcome
except through  treatment.  By convention, these  three conditions are
referred to as the exogeneity, monotonicity, and exclusion restriction
assumptions,  respectively  \citep{Angrist1996}.    When  a  valid  IV
exists, it  can be used to  draw inference about  the causal treatment
effect, despite the existence of unmeasured confounding.  However, the
IV  estimate  of the  treatment  effect  applies  only to  a  specific
non-identifiable subpopulation unless  additional assumptions are made
\citep{Angrist1996,  Imbens1994}.  In  simple settings  with  a binary
treatment  and  a binary  outcome,  IV methods  also  can  be used  to
construct   bounds   on   a   population   causal   treatment   effect
\citep{Robins1989, Manski1990, Joffe2001, Cheng2006, Zhang2003}, where
the uncertainty  of the impact of unmeasured  confounding is accounted
for by  a bound  (in contrast to  a point) estimate.   With continuous
outcomes, meaningful bounds are  not straightforward to obtain because
the domain of the outcome is typically unrestricted.

In this paper,  we consider the case of having  an observational study
with a continuous  outcome and a valid IV.  We  propose to combine the
IV and subjective assumptions, in  the context of the HIV Epidemiology
Research  Study  (HERS)  \citep{Smith1997},  to  construct  meaningful
bounds on  (1) the population  average treatment effect (ATE)  and (2)
the degree of unmeasured confounding.  The HERS was conducted when the
highly active antiretroviral therapy (HAART) first became available to
HIV-infected patients  but was  not randomly assigned.   Of particular
interest was  the initial-stage  causal effect  of HAART  on patient's
CD4+ T lymphocytes (CD4) count,  an important immunological marker for
immune system function and disease  stage.  The study was conducted at
two  types  of  study  site, community  clinics  and  academic  health
centers, which we use as an  IV.  The study had collected an extensive
set  of  covariates which  can  be  used to  adjust  for  part of  the
confounding effect, but unmeasured confounding may still exist and the
magnitude of  its impact is unclear.   To have a sense  of its impact,
many HIV-positive individuals in the early HAART era were reluctant to
initiate therapy due to fear of adverse side effects and toxicity.  At
the same time,  physicians tended to prescribe HAART  to patients with
poor  health  condition,  particularly  with  low  CD4  count.   These
confounding  factors  were  not  fully  measured  and  could  possibly
confound the HAART effect in a non-negligible way.

To  account for  the  measured confounding  effect,  we implement  the
inverse probability  weighting method proposed  by \citet{Robins1994},
assuming  that each  subject  has a  probability between  0  and 1  to
receive  HAART.  In  the  ideal case  when  unmeasured confounding  is
absent, the IPW  method can consistently estimate the ATE,  and can be
augmented to achieve  double robustness \citep[see][]{Bang2005}.  With
unmeasured confounding, we adopt the method of \citet{Robins:1999} and
incorporate a sensitivity parameter  into the IPW estimating equations
to  capture the  effect  of unmeasured  confounding.  The  sensitivity
parameter, defined  as the  systematic difference between  the treated
and untreated patients if hypothetically having these patients exposed
to the same  treatment condition, provides a measure  of the magnitude
of unmeasured confounding.  Without external information, however, the
sensitivity  parameter is  not  identified by  observed  data.  So  in
practice,  this  parameter is  often  used  to conduct  a  sensitivity
analyses to  assess the robustness  of the estimated  causal treatment
effect \citep{Ko2003,Brumback2004} to unmeasured confounding.  

In this paper, differential HAART  prescription rates at the two types
of study  site (study site  used as an IV)  provide an extra  piece of
information  that indeed  allows  us  to infer  the  magnitude of  the
sensitivity parameter.  In  this paper, we impose a  set of subjective
assumptions in the context of  the HERS.  These subjective assumptions
are  encoded  in  terms  of  the  potential  outcomes  as  well  as  a
sensitivity  parameter for  unmeasured confounding.   Putting together
the sensitivity parameter, the IPW estimating equations to account for
measured confounding, and the IV estimating equation leads to a system
of estimating  equations, unified  under a  constraint imposed  by the
principal  stratification   \citep{Frangakis2002}.   By   solving  the
unified equations, we  achieve our two objectives to  (1) estimate the
population average treatment effect of  HAART at the initial treatment
stage and  (2) quantify  the magnitude  of unmeasured  confounding.

The rest of the paper is  organized as follows: More details about the
HERS are provided in Section~\ref{sec:HERS}.  Notations and models are
given       in      Section~\ref{sec:Definitions-and-Model}.        In
Section~\ref{section:review.causal.methods}, we  review the  IV method
and  the IPW  method, and  introduce  a unified  system of  estimating
equations based on them.   In Section~\ref{sec:unified.ee}, we present
a  set  of  contextually  plausible constraints  and  assumptions  and
develop bounds on  the average treatment effect of HAART  on CD4 count
and    on    the    degree     of    unmeasured    confounding.     In
Section~\ref{section:application},  we  analyze  the  HERS  data,  and
finally in Section~\ref{section:discussion}, we  offer some points for
discussion.

\section{Motivating Example: HERS} 
\label{sec:HERS}

The  HERS was  conducted  from 1993-2001  to  investigate the  natural
history of  HIV progression in women.   Details of the HERS  have been
reported  in  \citet{Smith1997}.   In  the   study,  a  total  of  871
HIV-infected women were  enrolled at four study  sites: Baltimore, New
York  City,  Detroit,  and  Providence.   The  first  two  sites  were
community clinics while  the other two were  academic medical centers.
Clinical  outcomes such  as CD4  count were  recorded about  every six
months  since  enrollment.   Starting  from  1996,  HAART  became  the
recommended treatment regimen for  HIV infected people, especially for
those  with low  CD4 counts  \citep{Carpenter2000}.  Our  analyses use
data  extracted from  201  women  at their  8th  visit,  who 6  months
previously   were  HAART-naive   and  had   low  CD4   counts  ($<350$
cells/mm$^{3}$).   Using  the  HERS  data, we  want  to  estimate  the
initial-stage  causal  effect  of  HAART   on  CD4  count  among  this
population.  The  study had  collected a rich  set of  covariates, but
unmeasured confounding might still exist.

The   characteristics   of   the   201   women   are   summarized   in
Table~\ref{table:desc}.   Among them,  46  (23\%)  have initiated  the
HAART.  Those receiving HAART have a  higher CD4 count on average than
those  not  on  HAART,   but  this  ``as-received''  treatment  effect
\citep{Ten-Have:2008}  is  not   statistically  significant  (standard
normal z statistic = 0.58).  \citet{Ko2003} analyzed the data from the
HERS and screened out several  candidate confounders, which we list in
the  upper  panel  of   Table~\ref{table:desc}.   In  brief,  patients
receiving HAART  are more likely to  be aware their HIV  status and on
HAART at enrollment and at the previous visit; less likely to show any
HIV symptom and  be a drug user; have higher  viral loads (HIV-RNA) at
enrollment and at  the previous visit; and consist  of relatively more
white and less black.

In this paper, the  type of study site is used as  an IV assuming that
conventional IV assumptions  (outlined in Section \ref{sec:notations})
are  satisfied.  The  validity  of making  these  assumptions will  be
discussed in Section \ref{section:discussion}.   In the lower panel of
Table~\ref{table:desc}, we summarize the patient's CD4 count and HAART
receipt rates stratified by the type of study site.  Notably, patients
at academic centers are more likely  to be prescribed HAART (28 versus
18\%) than those at the community clinics, and their average CD4 count
is slightly  higher.  With the exogeneity  assumption, this difference
in CD4  count is the  causal effect of  study site.  Further  with the
exclusion restriction,  it is  the causal  effect of  the differential
HAART  assignment  between the  two  types  of  study sites.   In  the
following, we  will explore using  this extra piece of  information in
conjunction with other assumptions to infer the causal effect of HAART
and unmeasured confounding.

\begin{table}[!p]
  \caption{Summary of patient demographic characteristics by HAART
    receipt status and study site.  The numbers inside parentheses are
    standard errors. z stands for a standard normal test for comparing two sample means, and $\chi_{2}^{2}$ for a chi-squared
    statistic for Pearson's chi-squared test.}
  \label{table:desc} 
  \begin{tabular}{lccc}
    \toprule 
    & \multicolumn{2}{c}{Received HAART?} & Comparison \\
    \cmidrule{2-3} 
    & Yes  & No  & statistic \\
    \midrule 
    Number of patients, n  & 46  & 155  & - \\
    Average CD4 counts, cell/mm$^{3}$ & 229 (19)  & 216 (11)  & z $=0.58$ \\
    \emph{Candidate confounders } &  &  & \\
    ~~Race:  &  &  & \\
    ~~~~~ black; white; others  & 46; 28; 26\%  & 61; 15; 24\%  & $\chi_{2}^{2}=5.1$\\
    ~~ART receipt rate  &  &  & \\
    ~~~~~ at enrollment, \%  & 50\% (7.4)  & 39\% (3.9)  & z $=1.2$ \\
    ~~~~~ at previous visit, \%  & 74\% (6.5)  & 57\% (4.0)  & z $=1.9$ \\
    ~~Presence of HIV symptom, \%  & 26\% (6.5)  & 37\% (3.9)  & z $=1.2$ \\
    ~~HIV RNA, log$_{10}$ copy/mm$^{3}$  &  &  & \\
    ~~~~~ at enrollment, average  & 3.2 (.15)  & 3.1 (.07)  & z $=.78$ \\
    ~~~~~ at previous visit, average  & 3.7 (.15)  & 3.4 (.09)  &  z $=1.5$ \\
    ~~Intravenous drug use  &  &  & \\
    ~~~~~ recent, \%  & 22\% (6.1)  & .25 (.035)  & z $=.19$ \\
    ~~~~~ lifetime, \%  & 61\% (7.2)  & .63 (.039)  & z $=.04$\\
    ~~Aware of HIV status \%  & 83\% (5.6)  & .81 (.032)  & z $=.08$ \\
    \midrule
    & \multicolumn{2}{c}{The HERS study site} & \\
    \cmidrule{2-3} 
    & Academic centers  & Community clinics  & \\
    \midrule
    Number of patients, n  & 93  & 108  & - \\
    HAART received, n; \%  & 26; 28\%  & 20; 18\%  & z $=1.4$\\
    Average CD4, cell/mm$^{3}$  & 230 (14)  & 210 (12)  & z $=1.0$ \\
    \bottomrule 
\end{tabular}
\end{table}

\section{Notations and Definitions}
\label{sec:Definitions-and-Model}

\subsection{Notations}
\label{sec:notations}

We use $Z$ to  denote the IV (in the HERS, $Z=1$ if  the study site is
an  academic  medical center  and  $=0$  otherwise), and  $A_{z}$  the
potential treatment status that an individual would receive should $Z$
be  set to  $z$.   Hence,  each individual  has  a  pair of  potential
treatments $(A_{1},A_{0})$  that she would potentially  receive at the
two  types of  study site.   The \emph{actual}  treatment received  is
$A=A_{Z}=A_{1}Z +  A_{0}(1-Z)$, where $A=1$ means  that the individual
receives HAART,  and $0$  otherwise.  With  the Stable  Unit Treatment
Value Assumption~\citep{Rubin1974},  we use  $Y_{z}(a)$ to  denote the
potential outcome for  an individual should we  hypothetically set the
IV to $z$  and her treatment to $a$.  Thus,  the \emph{actual} outcome
observed  is   $Y=Y_{Z}(A)=Y_{Z}(A_{Z})$.   Further,  we   denote  all
confounders by  a vector $X$, and  the measured confounders by  $V$, a
subvector of  $X$. The  observed data consist  of $n$  identically and
independently  distributed  copies  of  $\{X_{i},Z_{i},A_{i},Y_{i}\}$,
$i=1,2,\ldots,n$.

We   assume   that   the    conventional   IV   assumptions   --   the
\emph{exogeneity},       \emph{exclusion       restriction},       and
\emph{monotonicity} assumptions  \citep{Angrist1996,Imbens1994} -- are
satisfied.     The   exclusion    restriction   assumes    $Y(a)\equiv
Y_{1}(a)=Y_{0}(a)$, i.e.\ the  IV has no direct effect  on the outcome
beyond  its  impact  on   individual's  treatment.   The  monotonicity
assumption requires  $\Pr(A_{1}\ge A_{0})=1$, i.e.\ an  individual who
would not receive HAART at an  academic medical center would not do so
at  a community  clinic either.   The exogeneity  assumes that  $Z$ is
jointly  independent   of  the  potential  outcomes   and  treatments,
$Z\perp(A_{0}, A_{1}, Y(0), Y(1))$.

\subsection{Definitions of Causal Treatment Effect}

The  causal effect  of HAART  treatment  can be  defined at  different
levels.       The      average       treatment      effect      (ATE),
$\mathrm{E}\{Y(1)-Y(0)\}$, is defined over the entire population.  The
ATE is of broad interest in public health and epidemiology, and is the
parameter of  interest in this paper.   With an IV, the  local average
treatment effect  \citep[LATE; ][]{Imbens1994}, $\mathrm{E}\{Y(1)-Y(0)
\mid A_0=0, A_1=1\}$, is defined as the average treatment effect among
a subpopulation who would receive  the treatment only when $Z=1$.  The
LATE can be estimated given a valid IV, not subject to the presence of
unmeasured confounding.  However, the facts that this subpopulation is
not fully identifiable  and the interpretation of the  LATE depends on
the  choice  of IV  pose  a  significant limitation  for  generalizing
results to a broader population and to other settings.

The relationship between  the ATE and LATE can  be expressed using the
principal   stratification   \citep{Frangakis2002}.    For  a   binary
instrument  and  a  binary  treatment,  the  principal  stratification
suggests  that the population  can be  partitioned into  four mutually
exclusive  subpopulations  based  on  the  potential  treatments  each
individual would have.  In our case, the potential treatments have the
following  four  possible  combinations, $(A_{0},A_{1})  \in  \{(0,0),
(0,1),  (1,0), (1,1)\}$,  where $\{(A_0,  A_1)=(0,0)\}$  indicates the
subpopulation   who   would   never   receive  the   HAART;   $\{(A_0,
A_1)=(0,1)\}$ is the subpopulation who would receive HAART only at
academic medical centers; and  so forth.  The monotonicity assumption,
$\Pr(A_1\ge   A_0)=1$,  implies   that   the  subpopulation   $\{(A_0,
A_1)=(1,0)\}$ is  an empty set.   Henceforth, we denote  the remaining
three  subpopulations  by  $\mathcal{P}_{jk}  =\{A_{0}=j,  A_{1}=k\}$,
$j\le k$.

We   denote   the  ATE   and   LATE   by  $\beta^{\mathrm{ATE}}$   and
$\beta^{\mathrm{LATE}}$, respectively.  Then their relationship can be
expressed by
\begin{equation}
  \label{equation:pr.str}
  \beta^{\mathrm{ATE}}=\pi_{01}\beta^{\mathrm{LATE}}+\pi_{00} 
  \{\mu_{00}(1)-\mu_{00}(0)\}+\pi_{11}\{\mu_{11}(1)-\mu_{11}(0)\}.
\end{equation}
where $\pi_{jk}=\Pr(\mathcal{P}_{jk})$  and $\mu_{jk}(a)  = \mathrm{E}
\{Y(a) \mid \mathcal{P}_{jk}\}$.  In  this paper, this relationship is
used to unify the IPW and IV  estimation methods, and based on that, a
system of estimating  equations is developed to draw  inference on the
ATE of HAART as well as the magnitude of unmeasured confounding.

\section{Review of Estimation Methods}
\label{section:review.causal.methods}

We review the IPW and IV methods in this section. 

\subsection{The IPW Method}
\label{section:IPW}

Putting aside  the covariates for  the moment, the  potential outcomes
can be expressed by a marginal structural mean model~\citep{Gange2007,
  Robins1999, Robins2000}
\begin{equation}
  \label{equation:strct.model}
  \mathrm{E}[Y(a)]=\beta_{0}+\beta^{\mathrm{ATE}}a,\quad 
  a=0,1.
\end{equation}

Assuming that $Y(a)\perp A|V$, i.e.\ unmeasured confounding is absent,
we    can   estimate    $\beta^{\mathrm{ATE}}$    by   the    solution
$\hat{\beta}_{\mathrm{IPW}}$ to the IPW estimating equations
\begin{equation*}
  U_{1}(\beta_{\mathrm{IPW}}) := \sum_{i=1}^{n} (1,A_{i})^{\top} W_{1i}
  (Y_{i}-\beta_{1}-A_{i} \beta_{\mathrm{IPW}})=0 , 
\end{equation*}
where  $W_{1i}={A_{i}/ \propscore(V_{i};  \gamma)} +  {(1-A_{i})/ \{1-
  \propscore(V_{i}; \gamma)\}}$,  and $e(V;\gamma)=\Pr(A=1|V)$  is the
propensity score \citep{Rosenbaum1983}  with a $l$-dimension parameter
$\gamma$.  We assume that $0<e(V;\gamma)<1$.

The IPW method has several  properties that are worth mentioning.  The
efficiency  of  the  resulting  estimator can  be  improved  by  using
stabilized  weights  to   replace  $W_{1i}$  \citep{Hernan2001}.   The
estimator can be augmented to  achieve double robustness if we further
specify an outcome regression model on $Y$~\citep{Bang2005}. Moreover,
if   $\gamma$   is   unknown,   $\hat{\beta}_{\mathrm{IPW}}$   remains
consistent  when  $\gamma$  is  replaced  by  a  consistent  estimator
$\hat{\gamma}$ that solves
\begin{equation*}
  U_{2}(\gamma):=\sum_{i=1}^{n}W_{2i}\{A_{i}-\propscore(V_{i};
  \gamma)\}=0.
\end{equation*}
where  $W_{2i}$ is  an appropriate  weight function;  e.g.\ $W_{2i}  =
{\partial\propscore(V_{i};  \gamma)  /  \partial\gamma}$  in  logistic
regressions.

When unmeasured  confounding exists,  $U_{1}(\beta_{\mathrm{IPW}})$ is
biased, i.e.\  $E\{U_{1}(\beta_{\mathrm{IPW}})\}\ne0$.  In  this case,
\citet{Robins1999}  proposed  to  introduce  a  sensitivity  parameter
$\tau$  and  then  estimate  $\beta^{\mathrm{ATE}}$  by  the  solution
$\hat{\beta}_{\MIPW}(\tau)$ to  the following modified  IPW estimating
equations
\begin{equation*}
  \label{eq:1}
  U_{3}(\beta_{\MIPW},\tau) :=
  \sum_{i=1}^{n}(1,A_{i})^{\top} W_{1i}\{Y_{i}^{*} -
  \beta_{2}-A_{i}\beta_{\MIPW}\}=0
\end{equation*}
where $Y_{i}^{*} = Y_{i}  - \tau\{A_{i} - \propscore(V_{i}; \gamma)\}$
is the ``outcome'' corrected for  the selection bias due to unmeasured
confounding.  For a binary treatment as in this paper, the sensitivity
parameter can  be defined  as the contrast  of the  potential outcomes
between the treated and untreated conditional on $V$.
\begin{equation*}
  \tau=(a-a')\left[\mathrm{E}\{Y(a)|A=a,V\}-\mathrm{E}
    \{Y(a)|A=a',V\}\right]
\end{equation*}
with $a=1-a'$.   In the context of  the HERS, $\tau>0$ means  that the
HAART is preferentially given to those with higher CD4 counterfactuals
$Y(a)$; $\tau<0$  means the  opposite is true;  and when  $\tau=0$, no
unmeasured  confounding   is  implied  and  the   resulting  estimator
$\hat{\beta}_{\MIPW}(0) = \hat{\beta}_{\mathrm{IPW}}$.

Without additional information from data,  the parameter $\tau$ is not
identified.   Hence,  the   resulting  estimator  $\hat{\beta}_{\MIPW}
(\tau)$ is typically used to conduct a sensitivity analysis.  That is,
estimate $\beta^{\mathrm{ATE}}$ using  $\hat{\beta}_{\MIPW} (\tau)$ as
if   $\tau$  is   known,   and  then   examine   the  sensitivity   of
$\hat{\beta}_{\MIPW}(\tau)$ by  varying the  value of $\tau$  over its
plausible range \citep{Ko2003,Brumback2004}.

With an  IV and information  extracted by  IV, it becomes  possible to
draw inference  about the ATE as  well as $\tau$ which  quantifies the
degree of unmeasured confounding.

\subsection{The IV Method}

The  IV  methods  have  been   widely  used  in  econometric  research
\citep[c.f.][]{Wooldrige2002}.   In our  just-identified  case with  a
single binary  IV and a  binary treatment, the standard  IV estimating
equations are
\begin{equation*}
  U_{4}(\beta_{\mathrm{IV}}):=\sum_{i=1}^{n}(1,Z_{i})^{\top}
  (Y_{i}-\beta_{3}-\beta_{\mathrm{IV}}A_{i})=0.
\end{equation*}
Under the IV assumptions and $\mathrm{Cov}(Z,A)\ne0$, the solution
\begin{equation}
  \label{equation:IV.estimator}
  \hat{\beta}_{\mathrm{IV}}=\frac{\overline{YZ}/\bar{Z}- 
    \overline{Y(1-Z)}/\overline{(1-Z)}}{\overline{AZ}/\bar{Z}-
    \overline{A(1-Z)}/\overline{(1-Z)}}
\end{equation}
is    consistent   for    $\beta^{\mathrm{LATE}}$   \citep{Imbens1994,
  Angrist1996,Hernan2006}.  The  bars in \eqref{equation:IV.estimator}
calculate sample averages, e.g.\ $\overline{YZ}=\sum_{i}Y_{i}Z_{i}/n$.
One    important    property    of    the   IV    method    is    that
$\hat{\beta}_{\mathrm{IV}}$   remains   consistent  despite   of   the
existence of unmeasured confounding.

Under  the framework  of the  generalized  method of  moments, the  IV
estimating  equations are  solved  using the  two-stage least  squares
method  \citep{Angrist1995},  and  can further  incorporate  a  weight
matrix to allow  for heteroskedastic or correlated  residuals.  The IV
methods  also  can  be  generalized  to deal  with  multiple  IVs  and
non-continuous outcomes \citep[c.f.][]{Wooldrige2002}.

\section{A Unified System of Estimating Equations}
\label{sec:unified.ee}

With a set of covariates and an  IV in the HERS, we propose to combine
the IV  and IPW methods, and  develop a unified  system of estimation
equations  as  follows,
\begin{equation}
  \label{equation:eq.sys}
  (U_{2}(\gamma),U_{3}(\beta_{\MIPW},\tau),U_{4}
  (\beta_{\mathrm{IV}}))^{\top}=0,
\end{equation}
with   a  constraint   of  \eqref{equation:pr.str}.    Note  that   in
\eqref{equation:pr.str},  parameters  $\mu_{11}(0)$ and  $\mu_{00}(1)$
are the averages of unobserved  potential outcomes and not identified.
All  other parameters  are identified  because $\pi_{11}  = \mathrm{E}
(A=1|Z=0)$, $\pi_{00} = \mathrm{E}  (A=0|Z=1)$, $\pi_{01} = 1-\pi_{00}
- \pi_{11}$, $\mu_{11}(1)  = \mathrm{E}(Y|A=1,Z=0)$,  and $\mu_{00}(0)
=\mathrm{E}  (Y|A=0,Z=1)$  \citep[and  Rejoinder]{Angrist1996}.   When
natural  limits exist  on $\mu_{11}(0)$  and $\mu_{00}(1)$  e.g.\ with
binary outcomes, both  the ATE and $\tau$ are  partially identified to
bounds.  When no  natural limits exist as is our  case with continuous
outcomes,  additional prior  information  is needed  to implement  the
constrained estimating equations system, which we will discuss next.

We present three sets of assumptions in the context of the HERS.  Each
allows us  to identify  bounds on the  ATE and  unmeasured confounding
parameter     $\tau$.       In     Sections~\ref{section:bound1}     -
\ref{section:bound3},  we   assume  that   the  sample  size   $n$  is
sufficiently large such that the sampling variations of the estimating
equations    \eqref{equation:eq.sys}   is    ignored.    In    Section
\ref{section:finitN}  and \ref{section:UR},  we discuss  inferences on
the sampling uncertainty of bound estimates for a finite $n$.

\subsection{Assumption  on  the  Upper  Limits  of  $\mu_{11}(0)$  and
  $\mu_{00}(1)$}
\label{section:bound1}

The  outcome  variable   of  our  interest  is  CD4   count,  so  both
$\mu_{00}(1)$ and  $\mu_{11}(0)$ must  be greater  than zero.   In our
first set of assumptions, we make a simple assumption that there exist
two upper bounds that

\noindent   \textbf{Assumption    (A)}:   $0\le\mu_{00}(1)\le\xi_{1}$,
$0\le\mu_{11}(0)\le\xi_{0}$, with known $\xi_{0}$ and $\xi_{1}$.

Assumption (A) leads to a simplified version of the Robins-Manski type
bound  on the  ATE~\citep{Robins1989, Manski1990,  Zhang2003}.  It  is
straightforward to show that the ATE falls within the interval
\begin{equation*}
  [\BB(\xi_{0},0),\BB(0,\xi_{1})],
\end{equation*}
where    to    emphasize     the    unidentifiable    parameters    in
\eqref{equation:pr.str},  we define  $\BB(\mu_{11}(0), \mu_{00}(1))  =
\pi_{01} \times \mathrm{LATE}  + \pi_{11}\{\mu_{11}(1) - \mu_{11}(0)\}
+\pi_{00}\{\mu_{00}(1) -\mu_{00}(0)\}$.

Then the  bound on  $\tau$ can  be inferred by  finding the  values of
$\tau$     such     that     the    corresponding     solutions     to
$\hat{\beta}^{\MIPW}(\tau)$  are consistent  with the  above bound  on
ATE.    For   a   given  $\beta^{\mathrm{ATE}}$,   the   solution   to
$U_{3}(\beta_{\MIPW}, \tau)=0$ for $\tau$ is
\begin{equation*}
  \hat{\tau}_{n}(\beta^{\mathrm{ATE}},\gamma) =\frac{\overline{W_{1}A} 
    *\overline{W_{1}(Y-\beta^{\mathrm{ATE}}A)}-\bar{W}_{1}
    *\overline{W_{1}A(Y-\beta^{\mathrm{ATE}}A)}}{
    \overline{W_{1}A}*\overline{W_{1}(A-\propscore(V;\gamma))}
    -\bar{W}_{1}*\overline{W_{1}A(A-\propscore(V;\gamma))}}.
\end{equation*}
It    is    straightforward    to    verify    that    $\hat{\tau}_{n}
(\beta^{\mathrm{ATE}},  \gamma)$  is   a  non-increasing  function  of
$\beta^{\mathrm{ATE}}$, so the unmeasured confounding parameter $\tau$
is bounded by
\begin{equation*}
  [\TT(\BB(0,\xi_{1}),\gamma),\TT(\BB(\xi_{0},0),\gamma)],
\end{equation*}
where  $\TT(\beta^{\mathrm{ATE}},  \gamma)  \equiv  \hat{\TT}_{\infty}
(\beta^{\mathrm{ATE}}, \gamma)$.

\subsection{Constraint  on  Relationships  between  $\mu_{11}(0)$  and
  $\mu_{00}(1)$ and Identified Quantities}
\label{section:bound2}

Assumption (A) alone  is sufficient for identifying the  bounds on ATE
and $\tau$, but in practice the  two upper limits need to sufficiently
large and  the resulting  bounds can  be wide.   In the  following, we
consider  making assumptions  on  the relative  magnitude between  the
unidentifiable and identifiable quantities.

\noindent \textbf{Assumption (B)}: We assume that 
\begin{enumerate}
\item The average treatment effect among $\AT$ is no less than a known
  $\delta_{11}$,
  \begin{equation*}
    \mathrm{E}\{Y(1)-Y(0)|\AT\}=\mu_{11}(1)-\mu_{11}(0)\ge\delta_{11}.
  \end{equation*}
  A plausible  choice for  $\delta_{11}$ is zero,  that is,  we assume
  that on  average, $\AT$  who would  \emph{always} receive  HAART can
  \emph{on  average}  benefit from  HAART.   We  make this  assumption
  because although  suffering from  confounding bias, the  efficacy of
  HAART  on the  treated  patients has  been  demonstrated by  several
  contemporary studies.  Further, we impose a known lower bound on the
  average treatment effect among $\NT$ that
  \begin{equation*}
    \mathrm{E}\{Y(1)-Y(0)|\NT\}=\mu_{00}(1)-\mu_{00}(0)\ge\delta_{00}.
  \end{equation*}
  A negative value of $\delta_{00}$  implies that HAART can be harmful
  for those  who would  never receive HAART  at either  site.  Setting
  $\delta_{00}=0$ implies  that HAART is  also beneficial for  them on
  average.
\item The difference on $\mathrm{E}  \{Y(0)\}$ between $\AT$ and $\NT$
  is bounded above,
  \begin{equation*}
    \mathrm{E}\{Y(0)|\AT\}-\mathrm{E}\{Y(0)|\NT\}=\mu_{11}(0)
    -\mu_{00}(0)\le\delta_{y0}.
  \end{equation*}
  We can  set $\delta_{y0}=0$ by  our intuition that in  the untreated
  condition, people who would  \emph{always} receive HAART have higher
  degree of HIV  progression (lower CD4 on average,  compared to those
  who would \emph{never} receive HAART).
\item  The difference  of treatment  effects between  those who  would
  \emph{always} receive HAART and those who would \emph{never} receive
  HAART is bounded below,
  \begin{equation*}
    \mathrm{E}\{Y(1)-Y(0)|\AT\}-\mathrm{E}\{Y(1)-Y(0)|\NT\}=
    \{\mu_{11}(1)-\mu_{11}(0)\}-\{\mu_{00}(1)-\mu_{00}(0)\}
    \ge\delta_{trt}.
  \end{equation*}
  For  example, letting  $\delta_{trt}=0$ implies  that the  treatment
  effect on those who would always receive HAART is greater than those
  would never do so.
\end{enumerate}

Under this set of assumptions, we show that the ATE is bounded by
\begin{equation*}
  [\BB(c_{0},\mu_{00}(0)+\delta_{00}),~\BB(0,c_{1})]
\end{equation*}
and $\tau$ by 
\begin{equation*}
  [\TT(\BB(0,c_{1}),\gamma),~\TT(\BB(c_{0},\mu_{00}(0)+\delta_{00}),
  \gamma)],
\end{equation*}
where  $c_{0}   =  \min\{\mu_{11}(1)  - \delta_{11},   \mu_{00}(0)  +
\delta_{y0}\}$ and $c_{1} = \mu_{11}(1)+\mu_{00}(0)-\delta_{trt}$.

\subsection{Constraint Conditional on Measured Covariates}
\label{section:bound3}

For the HERS,  it may be more realistic to  assume that Assumption (B)
holds  conditional  on clinically  important  covariates  $V$.  So  we
propose our third set of assumptions as

\noindent \textbf{Assumption (B')}: 
\begin{enumerate}
\item  $\mathrm{E}   \{  Y(1)-Y(0)   |  \AT,V  \}   \ge  \delta_{11}$;
  $\mathrm{E} \{ Y(1)-Y(0) | \NT,V \} \ge \delta_{00}$.
\item $\mathrm{E}\{ Y(0)  | \AT,V \} - \mathrm{E}  \{Y(0)|\NT,V\} \le
  \delta_{y0}$.
\item   $\mathrm{E}\{    Y(1)-Y(0)   |    \AT,V   \}    -   \mathrm{E}
  \{Y(1)-Y(0)|\NT,V\}  \ge  \delta_{trt}$,  for  known  $\delta_{11}$,
  $\delta_{00}$, $\delta_{y0}$ and $\delta_{trt}$.
\item  Further,   we  assume  that  the   monotonicity  and  exclusion
  restriction   assumptions  hold   conditional   on   $V$,  and   the
  constraint~\eqref{equation:pr.str} becomes
  \begin{align}
    \label{equation:const2}
    \mathbf{\beta} & =  \pi*\beta^{\mathrm{LATE}}+\int_{\mathcal{V}}\
    \mathrm{E}\{Y(1)-Y(0)|\AT,V\}P(\AT|V)\mathrm{d} F(V)\nonumber \\
    & +\int_{\mathcal{V}}\mathrm{E}\{Y(1)-Y(0)|\NT,V\}P(\NT|V)\mathrm{d}
    F(V), 
  \end{align}
  where  $\mathcal{V}$  is the  support  of  $V$ with  a  distribution
  $F(V)$.   We write  the product  $\pi_{01}*\beta^{\mathrm{LATE}}$ as
  before  because   both  the   $\mathrm{LATE}$  and   $\pi_{01}$  are
  identified by the  data.  Again, no observed data  are available for
  $\mathrm{E}\{Y(0)|\AT,V\}$ and $\mathrm{E}\{Y(1)|\NT,V\}$, which are
  denoted by $\mu_{11}(0,V)$ and $\mu_{00}(1,V)$, respectively.
\end{enumerate}
Under  (B')   and  \eqref{equation:const2},  we  obtain   a  bound  on
$\mathrm{ATE}$
\begin{equation*}
  [\pi_{01}\times\mathrm{LATE}+\int_{\mathcal{V}}\BB_{V}(c_{0}(V), 
  c_{2}(V))\mathrm{d} F,~
  \pi_{01}\times\mathrm{LATE}+\int_{\mathcal{V}}\BB_{V}(0,c_{1}(V))
  \mathrm{d} F]
\end{equation*}
and a bound on $\tau$ 
\begin{equation*}
  [\TT(\pi_{01}\times\mathrm{LATE}+\int_{\mathcal{V}}\BB_{V}
  (0,c_{1}(V))\mathrm{d} F,\gamma),~
  \TT(\pi_{01}\times\mathrm{LATE}+\int_{\mathcal{V}}\BB_{V}
  (c_{0}(V),c_{2}(V))\mathrm{d} F,\gamma)],
\end{equation*}
where $\BB_{V}(\mu_{11}(0,V),\mu_{00}(1,V)) = [\mathrm{E}
\{Y(1)|\AT,V\} - \mu_{11}(0,V)]\Pr(\AT|V) + [\mu_{00}(1,V)
- \mathrm{E} \{Y(0)|\NT,V\}] \Pr(\NT|V)$, $c_{0}(V)
=\min(\mathrm{E}\{Y(1)|\AT,V\} -\delta_{11},
\mathrm{E}\{Y(0)|\NT,V\}+\delta_{y0})$, $c_{1}(V)
=\mathrm{E}\{Y(1)|\AT,V\} +\mathrm{E}\{Y(0)|\NT,V\} -\delta_{trt}$,
and $c_{2}(V)=\mathrm{E}\{Y(0)|\NT,V\}+\delta_{00}$.

\subsection{Inference from Finite Samples}
\label{section:finitN}

To  implement \eqref{equation:pr.str},  we can  assume two  regression
models $E(A|Z)  = \logit^{-1}  (\eta(Z; \theta_{1}))$ and  $E(Y|A,Z) =
\kappa(A,   Z;  \theta_{2})$   for  some   known  functions   $\eta(Z;
\theta_{1})$ and  $\kappa(A,Z; \theta_{2})$.  For binary  $Z$ and $A$,
we use  two saturated models  and specify that $\eta(Z;  \theta_{1}) =
\theta_{10} + \theta_{11}Z$ and $\kappa(A,Z; \theta_{2}) = \theta_{20}
+  \theta_{21}Z+  \theta_{23}A  + \theta_{23}AZ$  with  $\theta_{1}  =
(\theta_{10},   \theta_{11})^\top$   and  $\theta_{2}   =(\theta_{20},
\theta_{21},  \theta_{22},  \theta_{23})^{\top}$.  Here,  a  saturated
additive model  for $\kappa(A,Z;  \theta_{2})$ is compatible  with the
structural model (\ref{equation:strct.model}), since one can show that
\eqref{equation:strct.model}  suggests   that  $\mathrm{E}(Y|A,Z)$  is
linear in $A$, $Z$, and $AZ$.

Given  two  regression models,  we  can  obtain consistent  estimators
$\hat{\theta}_{1}$ and $\hat{\theta}_{2}$ by solving
\begin{equation*}
  U_{5}(\theta_{1},\theta_{2}):=\left(\begin{array}{c}
      \sum_{i=1}^{n}W_{3i}[A_{i}-\logit^{-1}\{\eta(Z_{i};\theta_{1})\}]\\
      \sum_{i=1}^{n}W_{4i}\{Y_{i}-\kappa(A,Z;\theta_{2})\}
    \end{array}\right)=0,
\end{equation*}
where $W_{3i}=\partial \logit^{-1} \{\eta(Z_{i}; \theta_{1})\} / \partial
\theta_{1}$ and $W_{4i} = (1, Z_{i}, A_{i}, A_{i}Z_{i})^{\top}$.
Then, we estimate $\hat{\pi}_{11} = \logit^{-1} \{\eta(0;
\hat{\theta}_{1})\}$, $\hat{\pi}_{00} = 1-\logit^{-1}
\{\eta(1;\hat{\theta}_{1})\}$, $\hat{\pi}_{01} = \logit^{-1} \{\eta(1;
\hat{\theta}_{1})\} - \logit^{-1} \{\eta(0; \hat{\theta}_{1})\}$,
$\hat{\mu}_{11}(1) = \kappa(1,0; \hat{\theta}_{2})$, and
$\hat{\mu}_{00}(0) = \kappa(0,1; \hat{\theta}_{2})$.

For Assumptions (A) and (B), the  function $\BB()$ can be estimated by
replacing   those  identifiable   quantities  with   their  consistent
estimators,    i.e.\   $\hat{\BB}    (\mu_{11}(0),   \mu_{00}(1))    =
\hat{\pi}_{01}     \hat{\beta}_{\mathrm{IV}}      +     \hat{\pi}_{11}
\{\hat{\mu}_{11}(1)  -  \mu_{11}(0)\}  +\hat{\pi}_{00}\{\mu_{00}(1)  -
\hat{\mu}_{00}(0)\}$.   Further, we  estimate $c_{0}$  by $\hat{c}_{0}
(\delta_{11}, \delta_{y0}) =  \min\{(\hat{\mu}_{11}(1) - \delta_{11}),
(\hat{\mu}_{00}(0)  +  \delta_{y0})\}$  and  $c_{1}$  by  $\hat{c}_{1}
(\delta_{trt})    =    \hat{\mu}_{11}(1)   +    \hat{\mu}_{00}(0)    -
\delta_{trt}$.  By substituting these estimates for their estimands in
the bounds,  we obtain  bound estimates  on the  ATE and  $\tau$.  The
results are summarized in Table~\ref{table:bound.estimation}.

To estimate  the bounds on  ATE and  $\tau$ under Assumption  (B'), we
proceed as follows:
\begin{enumerate}
\item [Step 1.] We assume two observed-data models conditional on $V$,
  $\mathrm{E}(A|Z,V)   =   \logit^{-1}(\eta(Z,V;  \theta_{3}))$,   and
  $\mathrm{E}(Y|Z,A,V)   =   \kappa(Z,A,V;  \theta_{4})$   for   known
  $\eta(Z,V;  \theta_{3})$   and  $\kappa(Z,A,V;   \theta_{4})$.   For
  example, we  can assume two  linear models without  interaction that
  $\eta(Z,V; \theta_{3}) = \theta_{30}  + \theta_{31}Z + \theta_{32}V$
  with       $\theta_{3}       =      (\theta_{30},       \theta_{31},
  \theta_{32}^{\top})^{\top}$   and   $\kappa(Z,A,V;   \theta_{4})   =
  \theta_{40}  + \theta_{41}Z  +  \theta_{42}A +  \theta_{43}^{\top}V$
  with   $\theta_{4}   =   (\theta_{40},   \theta_{41},   \theta_{42},
  \theta_{43}^{\top}    )^{\top}$.     Let   $\hat{\theta}_{4}$    and
  $\hat{\theta}_{3}$  denote   the  estimators  of   $\theta_{4}$  and
  $\theta_{3}$ by solving the corresponding estimating equations.
\item  [Step  2.]   With  the monotonicity  assumption  and  exclusion
  restriction    conditional     on    $V$,    we     estimate    that
  $\widehat{\mathrm{E}}    \{Y(0)|\NT,V\}    =    \widehat{\mathrm{E}}
  (Y|Z=1,A=0,V=V)       =       \mu(1,0,V;       \hat{\theta}_{4})$,
  $\widehat{\mathrm{E}}       \{Y(1)|\AT,V\}        =       \mu(0,1,V;
  \hat{\theta}_{4})$,      $\widehat{\Pr}(\AT|V)       =      \pi(0,V;
  \hat{\theta}_{3})$,    and   $\widehat{\Pr}(\NT|V)    =   1-\pi(1,V;
  \hat{\theta}_{3})$.     Then     we    estimate     the    functions
  $\hat{\BB}_{V}()$,     $\hat{c}_{0}()$,    $\hat{c}_{1}()$,     and
  $\hat{c}_{2}()$ by bringing in the above estimators.
\item [Step 3.]  We estimate  the distribution $F(V)$ by the empirical
  cumulative density function of $V$  and integrals by empirical sums,
  e.g.\   estimate  $\int_{\mathcal{V}}   \BB_{V}  (c_{0}(V),c_{2}(V))
  \mathrm{d} F$ by  $\sum_{i=1}^{n} \hat{\BB}_{V} (\hat{c}_{0}(V_{i}),
  \hat{c}_{2} (V_{i}))  / n$.  Then,  we substitute the  parameters in
  \eqref{equation:const2} by their estimates.
\end{enumerate}
The resulting bound estimates on the  ATE and $\tau$ are summarized in
Table~\ref{table:bound.estimation}.


\begin{table}[!p]
  \caption{Estimated bounds on ATE and $\tau$ under Assumptions (A), (B)
    and (B').} 
  \label{table:bound.estimation}
  \centering 
  \begin{tabular}{llc}
    \toprule
    & Parameter  & Estimated bound \\
    \midrule 
    (A)  & ATE  & $[\hat{\BB}(\xi_{0},0),~\hat{\BB}(0,\xi_{1})]$\\
    & $\tau$  & $[\hat{\TT}_{n}(\hat{\BB}(0,\xi_{1}),\hat{\gamma}),~\hat{\TT}_{n}(\hat{\BB}(\xi_{0},0),\hat{\gamma})]$ \\
    \midrule
    (B)  & ATE  & $[\hat{\BB}(\hat{c}_{0},\hat{\mu}_{00}(0)+\delta_{00}),~\hat{\BB}(0,\hat{c}_{1})]$ \\
    & $\tau$  & $[\hat{\TT}_{n}(\hat{\BB}(0,\hat{c}_{1}),\hat{\gamma}),~\hat{\TT}_{n}(\hat{\BB}(\hat{c}_{0},\hat{\mu}_{00}(0)+\delta_{00}),\hat{\gamma})]$ \\
    \midrule
    (B')  & ATE  &
    $[\hat{\pi}_{01}\hat{\beta}_{\mathrm{IV}}+\frac{\sum_{i}\hat{\BB}_{v}(\hat{c}_{0}(V_{i}),\hat{c}_{2}(V_{i}))}{n},
    \hat{\pi}_{01}\hat{\beta}_{\mathrm{IV}}+\frac{\sum_{i}u\hat{\BB}_{v}(0,\hat{c}_{1}(V_{i}))}{n}]$\\
    & $\tau$  &
    $[\hat{\tau}_{n}\{\hat{\pi}_{01}\hat{\beta}_{\mathrm{IV}}+\frac{\sum_{i}\hat{\BB}_{v}(0,\hat{c}_{1}(V_{i}))}{n}\},
    \hat{\tau}_{n}\{\hat{\pi}_{01}\hat{\beta}_{\mathrm{IV}}+\frac{\sum_{i}\hat{\BB}_{v}(\hat{c}_{0}(V_{i}),\hat{c}_{2}(V_{i}))}{n}\}]$ \\
    \bottomrule
  \end{tabular}
\end{table}

\subsection{Uncertainty Region for Estimated Bounds}
\label{section:UR}

An interval that provides  $(1-\theta)$100\% coverage probability on a
bound estimate is often called the $(1-\theta)$100\% \emph{Uncertainty
  Region} (UR)  to distinguish  it from a  confidence interval.   A UR
takes  into   account  both  the  sampling   variability  and  partial
identifiability.   Two types  of  URs are  considered  in this  paper,
\emph{point-wise} and \emph{strong} $(1-\theta)$100\% coverage URs.

A  point-wise UR  $(\hat{L},\hat{U})$  contains  any particular  value
$\varrho\in(L,U)$ with  a probability of at  least ($1-\theta$), where
$(L,  U)$  denotes the  true  bound  and  $\varrho$ is  the  parameter
generating the  data.  If the  $\hat{L}$ and $\hat{U}$  are consistent
estimates   and   asymptotically   normally   distributed   (CAN),   a
($1-\theta$)100\% point-wise UR is given by
\begin{equation*}
  {\rm UR_{P\hyph CAN}}=[\hat{L}-{\rm c^{*}}{\rm se}(\hat{L}), 
  \hat{U}+{\rm c^{*}}{\rm se}(\hat{U})],
\end{equation*}
where  se($\cdot$) is  the standard  error and  c$^{*}$ is  a critical
value.  When $U-L$ is large  compared to ${\rm se}(\hat{L})$ and ${\rm
  se}(\hat{U})$, c$^{*}$ can  be approximated by $\Phi^{-1}(1-\theta)$
where     $\Phi$     is      the     normal     cumulative     density
function~\citep{Vansteelandt2006}.

A strong  UR is defined  as an interval  that contains the  entire set
$(L,U)$    with   a    probability    of    at   least    ($1-\theta$)
\citep{Horowitz2000,   Vansteelandt2006}.   If   both  $\hat{L}$   and
$\hat{U}$ are CAN, a strong ($1-\theta$)100\% UR is
\begin{equation*}
  {\rm UR_{S\hyph CAN}}=[\hat{L}-{\rm c~se}(\hat{L}),\hat{U}+
  {\rm c~se}(\hat{U})],
\end{equation*}
with ${\rm c}=\Phi^{-1}(1-\theta/2)=1.96$.   Without assuming $\hat L$
and $\hat U$ to be CAN, a strong ($1-\theta$) UR can be obtained using
the    bootstrap   method.     Specifically,   let    $(\tilde{L}^{*},
\tilde{U}^{*})$ denote the estimated bound from a bootstrapped sample.
A bootstrap strong 95\% UR is the interval $(L^{*}, U^{*})$ satisfying
$\Pr^{*}(L^{*} \le \tilde{L}^{*}, \tilde{U}^{*} \le U^{*}) = 1-\theta$
and  $\Pr^{*}(\tilde{L}^{*}<L^{*})  = \Pr^{*}(\tilde{U}^{*}  >U^{*})$,
where $\Pr^{*}$ is the probability measure induced by the bootstrapped
resamples~\citep{Bickel1981}, and  so can  be obtained by  finding the
shortest  interval  ${\rm  UR_{S\hyph   BTS}}  =  (L^{*},U^{*})$  that
satisfies $\frac{\#(L^{*} \le \tilde{L}_{k}^{*} < \tilde{U}_{k}^{*}\le
  U^{*})}{K}\ge1-\theta$,   and  $\frac{\#(L^{*}\le\tilde{L}_{k}^{*})}
{K}  \simeq \frac{\#(U^{*}  \ge \tilde{U}_{k}^{*})}{K}$,  where $\#()$
counts the number of statements that hold for $k=1,2,\ldots,K$.

\section{HERS Data: Treatment Effect Estimation and Confounding Assessment}
\label{section:application}

\subsection{Preliminary Analyses}

The  upper  panel  of  Table~\ref{table:result1}  summarizes  the  IPW
estimate of the ATE and IV estimate of the LATE of HAART on CD4 count.
The IPW  uses the  variables listed  in Table~\ref{table:desc}  as the
measured confounders of $V$ and  assumes that $\propscore(V; \gamma) =
\logit^{-1}(\gamma^{\top}V)$.  The  IPW estimate  of the  ATE suggests
that  HAART can  boost patient's  CD4  count by  27 cells/mm$^{3}$  on
average  with a  95\%  confidence  interval of  ($-$16,  70).  The  IV
estimate of the  LATE suggests that for those who  would receive HAART
at academic  medical centers but  not at community clinics,  HAART can
increase CD4 count by 207 on average  with a 95\% CI of ($-$250, 664).
In  Table~\ref{table:result1}, we  also list  the ``as-treated''  (AT)
treatment effect,  which is estimated  by the contrast of  the average
CD4 counts between those actually  receiving HAART and those not.  The
difference between  the IPW and  AT estimates  can be regarded  as the
bias  of  AT estimate  that  is  attributable to  the  \emph{measured}
confounders.


\begin{table}
  \caption{Estimates of HAART treatment effect on CD4 and $\tau$. The 95\% confidence
    intervals (CI) for point estimates and 95\% uncertainty regions for
    bound estimates (${\rm UR_{P\hyph CAN}}$, ${\rm UR_{S\hyph CAN}}$,
    and ${\rm UR_{S\hyph BOOT}}$; see Section 5.5) are shown in \textbf{bold}
    font. We assume that $\xi_{0}=\xi_{1}=500$ for (A); and that $\delta_{00} =\delta_{11} 
    =\delta_{y0}=\delta_{trt}=0$
    for (B) and (B').} 
  \label{table:result1} 
  \centering 
  \begin{tabular}{llcc}
    \toprule
    &  & ATE  & $\tau$  \\
    \midrule
    {AT}  & Point estimate  & 13  & - \\
    & 95\% CI  & \textbf{$(-$30, 56)}  & \\
    {IPW}  & Point estimate  & 27  & - \\
    & 95\% CI  & \textbf{$(-$16, 70)}  & \\
    {IV}  & Point estimate  & 207  & - \\
    & 95\% CI  & \textbf{$(-$250, 664)}  & \\
    \midrule 
    Assumption: &  &  & \\
    (A):  & Bound estimate  & $(-$196, 256)  & $(-$229, 223) \\
    & $95\%~{\rm UR}_{{\rm P\hyph CAN}}$  & \textbf{$(-$229, 289)}  & \textbf{$(-$269, 266)} \\
    & $95\%~{\rm UR}_{{\rm S\hyph CAN}}$  & \textbf{$(-$235, 295)}  & \textbf{$(-$277, 274)} \\
    & $95\%~{\rm UR}_{{\rm S\hyph BTS}}$  & \textbf{$(-$233, 294)}  & \textbf{$(-$274, 273)} \\
    (B):  & Bound estimate  & $(20,231)$  & $(-$204, 7.5) \\
    & $95\%~{\rm UR}_{{\rm P\hyph CAN}}$  & \textbf{$(-$9, 280)}  & \textbf{$(-$260, 49)} \\
    & $95\%~{\rm UR}_{{\rm S\hyph CAN}}$  & \textbf{$(-$15, 289)}  & \textbf{$(-$271, 57)} \\
    & $95\%~{\rm UR}_{{\rm S\hyph BTS}}$  & \textbf{$(-$14, 285)}  & \textbf{$(-$270, 57)} \\
    (B'):  & Bound estimate  & $(18,218)$  & $(-$191, 9.1) \\
    & $95\%~{\rm UR}_{{\rm P\hyph CAN}}$  & \textbf{$(-$10, 261)}  & \textbf{$(-$234, 48)} \\
    & $95\%~{\rm UR}_{{\rm S\hyph CAN}}$  & \textbf{$(-$16, 270)}  & \textbf{$(-$243, 56)} \\
    & $95\%~{\rm UR}_{{\rm S\hyph BTS}}$  & \textbf{$(-$14, 270)}  & \textbf{$(-$243, 56)} \\
    \bottomrule
  \end{tabular}
\end{table}

\subsection{Bounds on HAART Treatment Effect and Unmeasured Confounding}

For  Assumption (A),  we let  the upper  limits $\xi_{0}=\xi_{1}=500$.
(Recall that the two limits are on the expected values of $Y(0)$ among
$\AT$ and $Y(1)$ among $\NT$.)  We  choose the two limits based on the
facts that the average CD4 count  at the previous visit was much lower
than 350 and  at the eighth visit,  the average CD4 count  was 229 for
those   treated    and   216    for   those   untreated    (refer   to
Tables~\ref{table:desc}).   For  Assumptions  (B)  and  (B'),  we  let
$\delta_{11}=\delta_{00}=\delta_{y0}=\delta_{trt}=0$, and  further for
(B') let $V$  be the variables listed  in Table~\ref{table:desc}.  The
lower  panel   of  Table~\ref{table:result1}  summarizes   the  bounds
estimates  of the  ATE  and $\tau$.   The bound  estimate  of the  ATE
$(-196,256)$ under  (A) is not  informative and much wider  than those
under  (B) (20,  231)  and (B')  (18, 218).   Assumption  (B') is  not
necessarily stronger  than (B), but  by imposing observed  data models
and  having the  estimated  bounds smoothed  over covariates,  tighter
bounds are resulted in.  To  obtain uncertainty regions on these bound
estimates, we draw  $K=1,000$ bootstrap samples, fixing  the number of
patients  at the  two types  of study  sites (academic  medial centers
versus  community  clinics).   Because the  bounds  estimates  contain
$\min()$ operation which complicates  the derivation of their standard
errors, we use the $K$  bootstrapped samples to calculate the standard
errors of the two  ends of bound estimates.  Table~\ref{table:result1}
summarizes the point-wise, strong,  and bootstrap strong 95\% coverage
URs.  The difference between the 95\%  CI of the IPW estimate and 95\%
URs under (B) and (B') can be regarded as the bias of the IPW estimate
due  to  \emph{unmeasured}  confounding.   The  results  suggest  that
unmeasured confounding tends to cause a downward bias and the true ATE
is likely to higher than what the IPW 95\% CI suggests.

The bound estimates on $\tau$ under  (B) and (B') are $(-204,7.5)$ and
$(-191,9.1)$, which  are much  tighter and  more informative  than the
bound estimate  under (A)  $(-229,223)$.  The 95\%  URs on  $\tau$ are
listed  in Table~\ref{table:result1}.   A possibly  negative value  of
$\tau$  implies  that  unmeasured  factors  resulted  in  preferential
prescriptions of HAART to those with  fewer CD4 count in the HERS, and
those  on  HAART  might have  up  to  200  fewer  CD4 count  (if  left
untreated) on average compared with those not on HAART.

\subsection{Sensitivity to Unknown Parameters}
\label{section:sens}

\begin{figure}[!p]
  \centering\includegraphics[width=0.8\textwidth]{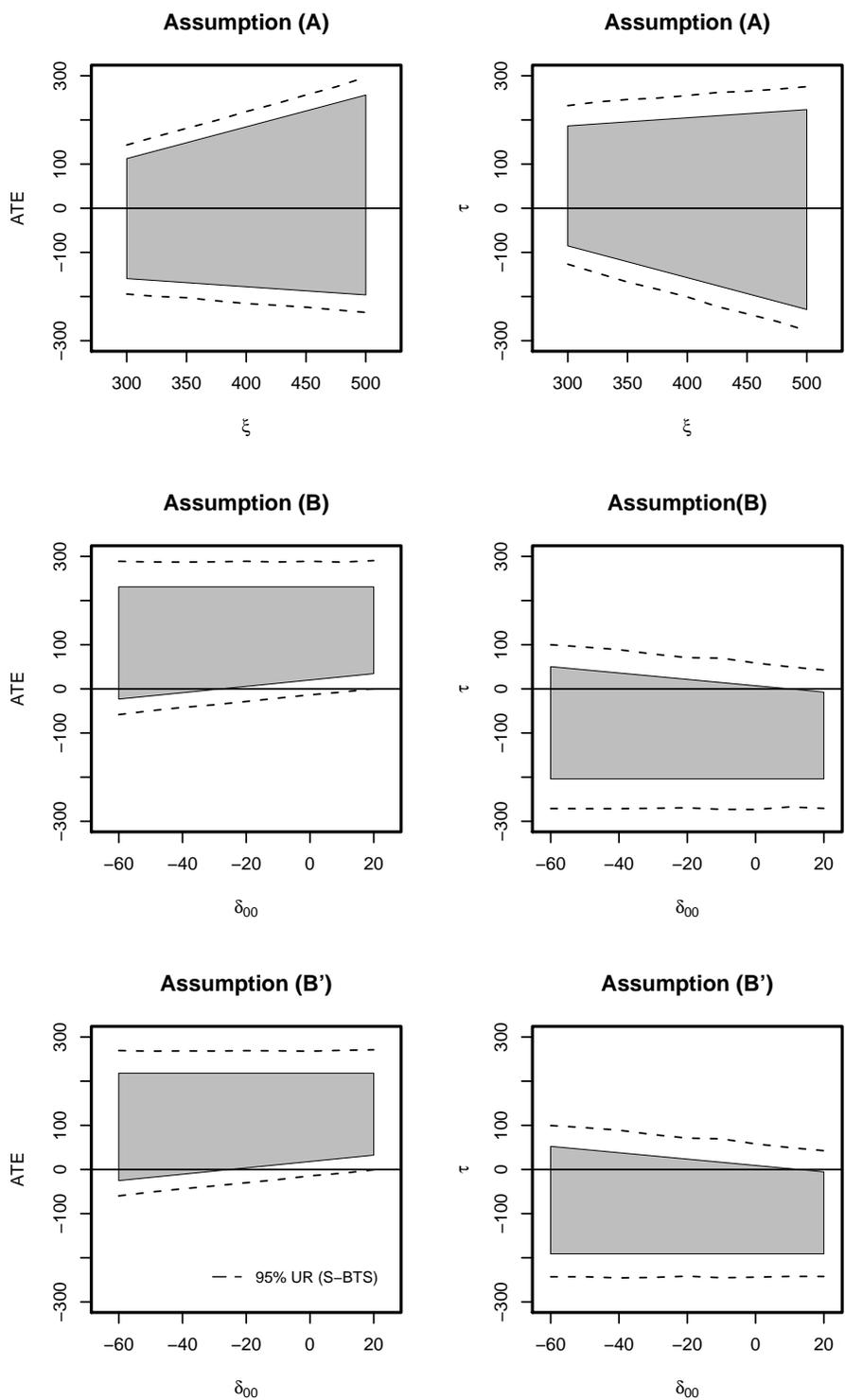}
  \caption{Sensitivities of bound estimates to $\xi= \xi_{0}=
    \xi_{1}$ under Assumption (A); to $\delta_{00}$ under Assumptions
    (B) and (B').  The gray zones show the bound estimates as a function of
    $\xi$ or $\delta_{00}$.  The bootstrap strong 95\% ${\rm
      UR_{S\hyph BTS}}$'s are shown as dashed lines.}
  \label{figure:sens}
\end{figure}

In  this section,  we conduct  a simple  sensitivity analysis  for the
unknown parameters used in the three sets of assumptions.  We impose a
common upper  limit $\xi=\xi_{0}=\xi_{1}$  for Assumption (A)  and let
$\xi$ vary  from $300$ to $500$.   The bound estimates on  the ATE and
$\tau$   along  with   bootstrap  strong   95\%  URs   are  shown   in
Figure~\ref{figure:sens} (First row).  For the considered range, $\xi$
has more  influence on  the upper  (lower) bound  estimate on  ATE (on
$\tau$),   and  the   resulting  bound   estimates  remain   wide  and
non-informative.


For (B) and  (B'), we let $\delta_{00}$ (the lower  limit of treatment
effect  among $\NT$)  rang from  $-60$  to $20$  and fix  $\delta_{11}
=\delta_{y0}  =  \delta_{trt}  =0$.  (More  sophisticated  sensitivity
analyses   that   jointly   evaluate   $\delta_{11}$,   $\delta_{00}$,
$\delta_{y0}$ and $\delta_{trt}$ are  possible.)  We choose this range
for $\delta_{00}$ based on the magnitude  of the AT and IPW estimates,
and have  it tilt toward  the negative  side for the  possibility that
HAART   could   be   harmful   for  those   never   receiving   HAART.
Figure~\ref{figure:sens}   (Second   and   third  rows)   shows   that
$\delta_{00}$ only  affects the lower  (upper) bound estimates  of the
ATE (of  $\tau$).  The ATE can  be as high as  over 200 cell/mm$^{3}$,
and the lower  bound of ATE varies around zero  depending on the value
of $\delta_{00}$.  Again, a possible negative value of $\tau$ suggests
that unmeasured  confounding likely causes HAART  to be preferentially
prescribed to those with poorer health.

\section{Discussions}
\label{section:discussion}

In the study of HERS, we propose to use an IV and sets of contextually
plausible assumptions to quantify the  causal effect of a treatment as
well as the degree of  unmeasured confounding.  We consider three sets
of assumptions.  Assumption  (A) specifies the limits  of the expected
unobservable potential  outcomes, which leads to  a simplified version
of the Robins-Manski bounds on  ATE.  Assumptions (B) and (B') specify
the relative magnitudes between  identified and unidentified potential
outcome averages,  and lead to  bounds that  can be much  tighter than
(A).  The 95\% uncertainty regions for  the ATE under (B) and (B') are
more informative  than the 95\% CI  of the IV estimate,  and have less
concern of having unmeasured confounding  bias compared to the 95\% CI
of the IPW estimate.  The bound estimates on the ATE and $\tau$ reveal
that unmeasured  confounding could  cause a downward  bias on  the ATE
because of HAART being preferentially  prescribed to those with poorer
health condition.

Quantifying the degree  of unmeasured confounding can  be valuable for
analysis of  studies conducted in  similar settings but having  no IV.
Several HIV observational studies~\citep[e.g. ][]{Gange2007} have been
conducted contemporarily  as the HERS,  and could suffer  from similar
amount of  unmeasured confounding.   In those studies  when unmeasured
confounding  is of  concern, analyses  should be  complemented with  a
sensitivity  analyses as  described in  Section~\ref{section:sens}.  A
plausible range for $\tau$ can be informed from our study.

In  this  paper, we  use  the  type of  study  site  as an  instrument
variable, assuming  that two crucial IV  assumptions (monotonicity and
exclusion restriction)  are satisfied.  The observed  HAART assignment
rate at  academic centers  is higher than  that at  community clinics.
Such  an  observation  suggests that  the  deterministic  monotonicity
$\Pr(A_{1}\ge A_{0})=1$  is plausible,  but this assumption  cannot be
verified.  As  one limitation  of our study,  this assumption  will be
violated if some individuals would  receive HAART at community clinics
but  not at  academic medical  centers.   If the  proportion of  these
individuals ($\DE$)  is small, the  bias due  to the violation  of the
monotonicity  assumption is  probably negligible.   Alternatively, one
can assume that $\NT$ is absent, so that $\AT$, $\CM$ and $\DE$ form a
partition of the population.  This  assumption allows everyone to have
some chance of  receiving HIV therapy, which is also  sensible for the
HERS because these  patients' CD4 counts are less than  350 six months
before,  and  allows  for  the  possibility  that  some  people  would
potentially  be treated  at a  community  clinic but  not an  academic
medical center.  With this assumption, the proportions of $\AT$, $\CM$
and $\DE$ are identified  because $\pi_{01} = \Pr(A=0|Z=0)$, $\Pr(\DE)
= \Pr(A=0|Z=1)$, and $\Pr(\AT)  = 1-\pi_{01}-\Pr(\DE)$.  The following
estimands   are   also    identified:   $\mathrm{E}   \{Y(0)|\CM\}   =
\mathrm{E}(Y|A=0,Z=0)$,  $\mathrm{E}   \{Y(0)|\DE\}  =  E(Y|A=0,Z=1)$,
$\mathrm{E}\{Y(1)|\AT  \textrm{or}  \CM\}=\mathrm{E}(Y|A=1,Z=1)$,  and
$\mathrm{E}\{Y(1)|\AT  \textrm{or}   \DE\}=\mathrm{E}(Y|A=1,Z=0)$.   A
challenge here is  how to incorporate the IV estimator,  which now has
an estimand as a `weighted'  contrast of the average treatment effects
between   $\CM$   and   $\DE$,   to   construct   constraint   similar
to~(\ref{equation:pr.str}).  This may be worth further investigation.

Moreover, replacing  the deterministic monotonicity with  a stochastic
monotonicity   assumption  deserves   explorations   in  the   future.
\citet{Roy2008} assumed  $\Pr(A_{1}=1 |  A_{0}=1,V) \ge  \Pr(A_{1}=1 |
A_{0}=0,V)$, and proposed to use  auxiliary covariates to estimate the
memberships   of   principal    strata.    \citet{Small2008}   assumed
$\Pr(A_{1}=1|U)\ge\Pr(A_{0}=1|U)$  with $U$  being  a latent  variable
satisfying   certain   conditions.   These   stochastic   monotonicity
assumptions  allow the  possible presence  of  $\DE$ and  may be  more
realistic in the HERS than the deterministic monotonicity.

The exclusion restriction could also be  violated if the type of study
site $Z$ remains associated with  the outcome $Y$ after accounting for
the effect  of $Z$ on  HAART receipt.  A weaker  exclusion restriction
assumption can be made, if  the association between the instrument and
the  outcome  can  be  removed after  conditioning  on  some  measured
covariate $V^{*}$, i.e.\ $\{Y(1),Y(0)\} \perp Z|V^{*}$.  In this case,
the methods by \citet{Tan2006} can  be implemented for identifying the
LATE, and  our method  for bound  estimation on  ATE and  $\tau$ still
applies.

There are  several ways to  account for the measured  confounding.  We
use  the  method of  inverse  probability  weighting by  specifying  a
propensity score model.  Alternatively, we can specify both an outcome
regression  model and  a propensity  score  model and  use the  doubly
robust (DR) estimator \citep{Bang2005} to estimate the ATE.  We do not
implement  the DR  estimator  in this  paper  because when  unmeasured
confounding  exists the  DR estimator  is no  longer guaranteed  to be
consistent for ATE  and could suffer more bias  than other estimators.
The  simulations of~\citet{Kang2007}  suggest that  IPW is  relatively
robust to the  impact of unmeasured confounding in  term of estimation
bias.   Because   the  focus  issue   of  this  paper   is  unmeasured
confounding, we use the IPW for estimating ATE.

\bibliographystyle{plainnat}
\bibliography{ivipw}

\end{document}